\documentclass[paper,notoc]{JHEP3}

\usepackage{cite}
\usepackage{epsfig}
\usepackage{amsfonts}
\usepackage{amsmath}
\usepackage{amssymb}

\def \beq {\begin{equation}}
\def \eeq {\end{equation}}
\def \bea {\begin{eqnarray}}
\def \eea {\end{eqnarray}}
\def \nn {\nonumber}

\def\Z#1{_{\lower2pt\hbox{$\scriptstyle#1$}}}
\def\X#1{_{\lower2pt\hbox{$\scriptscriptstyle#1$}}}

\title{Warped compactification on curved manifolds}

\author{Ishwaree P. Neupane\\
Department of Physics and Astronomy, University of Canterbury\\
Private Bag 4800, Christchurch 8020, New Zealand\\ and\\
Yukawa Institute for Theoretical Physics, Kyoto University\\ Kyoto
606-8502, Japan \\
\\ \email{E-mail:ishwaree.neupane@canterbury.ac.nz}}

\abstract{The characterization of a six- (or seven)-dimensional
internal manifold with metric as having positive, zero or negative
curvature is expected to be an important aspect of warped
compactifications in supergravity. In this context, Douglas and
Kallosh recently pointed out that a compact internal space with
negative curvature could help to construct four-dimensional de
Sitter solutions only if the extra dimensions are strongly warped
or there are large stringy corrections. That is, the problem of
finding 4-dimensional de Sitter solutions is well posed, if all
extra dimensions are physically compact, which is called a no-go
theorem. Here, we show that the above conclusion does not extend
to a general class of warped compactifications in classical
supergravity that allow a non-compact direction or cosmological
solutions for which the internal space is asymptotic to a cone
over a product of compact Einstein spaces or spheres. For clarity,
we present classical solutions that compactify higher-dimensional
spacetime to produce a Robertson--Walker universe with de
Sitter-type expansion plus one extra non-compact direction. Such
models are found to admit both an effective four-dimensional
Newton constant that remains finite and a normalizable zero-mode
graviton wavefunction. We also exhibit the possibility of
obtaining 4D de Sitter solutions by including the effect of fluxes
(p-form field strengths). }


\preprint{UOC-TP 012/10, YITP-10-48, arXiv:1006.4495}

\begin{document}

\section{Introduction}

Since the beginning of the 20th century, physicists have struggled
to find a way to unite general relativity and quantum mechanics.
One of the methods by which theoretical physicists have tried to
unify gravity with the other forces is to build models that allow
for many more spatial dimensions. A strong motivation for
considering additional dimensions of space came from string and
supergravity theories, which in turn are inspired by the failure
of classical gravity to work at very short distance scales. 

Given that our universe is described by fundamental theories of
higher-dimensional gravity, such as, string/M theory, there should
exist six or seven extra dimensions of space, not yet detected by
experiment. This is possible if the extra dimensions take the form
of a small compact manifold ${\cal M}$ or are strongly warped
along the transverse direction, giving rise to a finite warped
volume for these extra spaces.

In recent years, much of the focus has been on the study of
$(4+m)$-dimensional supergravity theories defined on
non-factorizable background spacetimes, instead of on the
conventional Kaluza-Klein (or product-space) compactifications,
$M^{m+4}\to M^4 \times M^m$. An explicit example of this category
is the five-dimensional braneworld model studied by Randall and
Sundrum~\cite{RS1}, which has brought new prospective in our
thinking about the role of gravity in higher dimensions, providing
an alternative to the standard Kaluza-Klein or product-space
compactification. The RS single brane model~\cite{RS2} has further
raised the possibility that, at least, one of the extra dimensions
postulated by string/M theory could be large enough to have some
new cosmological and phenomenological implications~\cite{Cline99}.

The 5D braneworld construction is sufficiently simple at a level
of model building in particle physics, but it is incomplete in
many ways. On general grounds, there is no strong reason to
consider just one extra dimension, instead of six or seven as
predicted by string or M theory. In the last ten years or so,  a
lot of work has been done in this direction, taking into account
the effect of background fluxes (form fields) and D-branes. The
existence of branes, fluxes and localized objects like an
orientifold plane in some fundamental theories of gravity,
including string theory, has benefited physicists to develop new
ideas, including the localization of gravity on D3-branes,
dualities between closed string theories that contain gravity and
decoupled open string (or gauge) theories~\cite{Malda97,KS} and
methods of constructing metastable de Sitter vacua using warped
extra dimensions~\cite{KKLT}.

Despite several novelties of string/M theory, including
microscopic descriptions of inflation from D-braneworld
models~\cite{Burgess-etal,Sami07,Becker07,Cassani09,Lust09}, it is
not straightforward to explain an accelerated expansion of the
universe directly through compactification of supergravity models
in 10 or 11 dimensions (which are the low energy limits of string
theory and M-theory). Specifically, there is a `no-go theorem',
due to Gibbons~\cite{Gibbons-84}, De Wit {\it et al}~\cite{WSH87},
Maldacena and Nunez~\cite{Malda-Nunez}, Giddings {\it et
al}~\cite{GKP} and many others, which basically asserts that if we
dimensionally compactify any string-derived supergravity model on
a smooth compact internal manifold ${\cal M}$, then we often end
up with a flat Minkowski space as a viable background solution of
classical supergravities, unless we violate certain positivity
conditions. The problem of finding de Sitter solutions is well
posed, if all extra dimensions are physically compact. Since the
universe is evidently both past and future de Sitter (albeit with
vastly differing vacuum energies) this would seem to be a problem.

In recent years, attempts have been made around this particular
`no-go' result. The original ``no-go" theorem assumes time
independence of the internal space, and so one could look for
time-dependent solutions. Following this intuition, Chen {\it et
al}~\cite{Chen:2002}, Townsend and Wohlfarth~\cite{TW03},
Ohta~\cite{Ohta03}, Neupane {\it et al}~\cite{Ish05} and
others~\cite{GKP} constructed time-dependent solutions of ten and
eleven-dimensional supergravities which describe one or more
periods of transient acceleration of the universe; we refer
to~\cite{Ish03,Maeda-all} for further references. The basic idea
is simple: given that the internal space is described by certain
metric moduli associated with the internal space scale factors and
these moduli describe the size and other basic properties of the
internal manifold, then upon dimensional reduction, time-dependent
metric moduli typically give rise to an exponential potential in
lower dimensions~\cite{Ish03a}. In the case of pure
$D$-dimensional gravity~\cite{Ish05,Ish03}, one would require one
or more internal subspaces to have negative curvature,
particularly, with time-dependent metric moduli, which generate a
sum of positive exponential potentials in lower dimensions. The
background fluxes provide an additional contribution to them. One
can now imagine `bouncing' the universe off an effective
potential~\cite{Emparan03}. Albeit for a brief interval, the
energy is dominated by the potential term, and the universe
undergoes a transient period of cosmic acceleration. All higher
dimensional models with time-dependent metric moduli field inherit
the property that as one moves to the minimum of an effective
potential the size of extra dimensions grows only slowly or even
stabilise in a few specific cases~\cite{Ish06}.

One particular reason to be interested in warped geometries (with
an internal space of arbitrary curvature) is to find 4D de Sitter
solutions. To this end, one finds natural interest in supergravity
models, where the extra dimensions are warped but
time-independent. In this context, Douglas and Kallosh recently
pointed out that a compact internal space with negative curvature
could help us to construct four-dimensional de Sitter solutions
only if the extra dimensions are strongly warped or there are
large stringy corrections~\cite{DK10}. That is, the problem of
finding 4-dimensional de Sitter solutions is well posed, if all
extra dimensions are physically compact, which is called a no-go
theorem. In this paper, we show that within a certain class of
warped supergravity models it won't be necessary to introduce
stringy corrections and/or some localized brane sources violating
certain positivity conditions just for the purpose of getting 4D
de Sitter solutions. Here we give some explicit examples for which
the 4D effective Planck mass is finite, despite having a
noncompact direction, and the 4D de Sitter solutions exist with an
arbitrary choice of the internal curvature.

The earlier results, for instance, in~\cite{KS,GKP} show that, in
a non-cosmological setting, a nontrivial warp factor could arise
as the backreaction of some (background) fluxes. However, in a
cosmological setting, in which one replaces a flat 3-brane by some
cosmologically relevant metrics, such as, FLRW universe, we find
that a non-trivial warp factor could by itself give rise to de
Sitter solutions in four dimensions. We show that one can obtain a
de Sitter solution even when the internal manifold is Ricci flat,
provided that the warp factors for both the external 4D spacetime
and the internal 6D space are nontrivial. This work significantly
extends and generalizes the earlier analysis in~\cite{Ish09a}.

Some of the examples of solutions presented in this paper may look
closer in spirit to Randall--Sundrum-type braneworld
constructions~\cite{Others}. The important difference here is that
the effect of a compact base manifold $X_5$ (or $X_6$ in the
$D=11$ case) is also incorporated in the classical solutions. Our
construction is motivated from 10D and 11D supergravity theories,
in which one takes into account the effect of $p$-form gauge
fields.

The paper is organized as follows: In section 2, we start with a
warped 10D metric ansatz, which is the product of a maximally
symmetric 4D spacetime and a general 6D Einstein space of
arbitrary curvature. We solve the 10D Einstein equations without
fluxes (or source terms) and show that they lead to de Sitter
solutions in four dimensions. Our examples lead to a finite warped
volume and hence a finite 4D effective Planck mass.

In section 3, we briefly discuss on why the `warping' of extra
spaces might play an important role in the construction of de
Sitter solutions, generalizing earlier works
in~\cite{Ish09a,Ish09b}. We also discuss the differences and
advantages of adding energy sources for p-branes and fluxes. In
section 4, we analyze 10D supergravity flux equations. We use our
knowledge of the background de Sitter solutions to explore the
possibility of having a negatively curved 6D space but again
without introducing brane sources or objects violating `internal
energy' conditions. The addition of $p$-form fluxes does not much
affect the nature of 4D de Sitter solutions except in the limit
where the base manifold $X_5$ shrinks to zero size. In section 5,
we present a four-dimensional de Sitter solution by solving
Einstein's equations in 11 dimensions. In section 6, we briefly
summarize our results.

\section{Warped metrics}


We begin with a ten-dimensional metric ansatz of the form
\begin{eqnarray}
ds_{10}^2 = \tau^2 \, e^{\beta A(y)} \,\hat{g}_{\mu\nu} dx^\mu
dx^\nu + \rho^2\, e^{\alpha A(y)} ds_6^2,\label{main-10d-anz}
\end{eqnarray}
with
\begin{equation}
ds_6^2= g^{(6)}_{ij} (y) dy^i dy^j,
\end{equation}
where $\beta, \alpha$ are numerical constants, and $\tau^2$ and
$\rho^2$ are some other constants which can be related to
extremized values of 10D dilaton and volume modulus. 
%
For the metric (\ref{main-10d-anz}), with $X_6$ as a compact
internal space, we find
\begin{equation}
{}^{(10)}R_{\mu\nu}(x,y) =
{}^{(4)}\hat{R}_{\mu\nu}(x)-\frac{\tau^2}{\rho^2}
(\beta^2+\beta\alpha) (\nabla A)^2 e^{(\beta-\alpha)A(y)}
\hat{g}_{\mu\nu}-\frac{\tau^2}{\rho^2} \frac{\beta}{2}
e^{(\beta-\alpha)A(y)} \hat{g}_{\mu\nu} \nabla_y^2
A,\label{symm-4D}
\end{equation}
\begin{eqnarray}
{}^{(10)}R_{ij}(x,y)&=& {}^{(6)}\tilde{R}_{ij}
-(\alpha^2+\beta\alpha) (\nabla A)^2 \tilde{g}_{ij}^{(6)}-
2(\beta+\alpha)\nabla_i \nabla_j A \nonumber \\
&{}& -(\beta^2-2\beta\alpha-\alpha^2)\nabla_i A \nabla_j A
 -\frac{\alpha}{2}\tilde{g}_{ij}^{(6)} \nabla_y^2 A.\label{symm-6D}
\end{eqnarray}
This result corrects a couple of typos/errors
in~\cite{Danielsson09}. In the above $\tilde{g}^{(6)}_{ij}$ denote
the metric components of the internal space, which are independent
of the $y$ coordinate. The notations here follow that from
refs~\cite{Ish09a,Ish09b}.

\subsection{A canonical choice of 6D metric}

The simplest examples of 6D compact spaces are given
by~\footnote{In the $\epsilon=-1$ case, a compact hyperbolic
manifold is achieved by taking a quotient $H^6/\Gamma|_{\rm free}$
of the non-compact space $H^6$ by a freely acting discrete
subgroup $\Gamma$ of the isometry group.}
\begin{subequations}
\begin{align}
ds\Z{6}^2 &= dy^2+ dy_1^2+ \cdots + dy_5^2,\quad
(\epsilon=0)\label{6d-flat}\\
ds\Z{6}^2 &= dy^2+ \sin^2{y}
\,d\Omega_{5}^2,\quad (\epsilon=+1) \label{6d-positive}\\
ds\Z{6}^2 & = dy^2+ \sinh^2{y}\, d\Omega_{5}^2,\quad
(\epsilon=-1), \label{6d-negative}
\end{align}
\end{subequations}
where $d\Omega_{5}^2$ represents the metric of a 5-sphere,
${}^{(6)}\tilde{R}_{ij}=\epsilon (m-1) \tilde{g}_{ij}$, with $m$
being the number of extra dimensions and $m=6$ in this example.
Furthermore,
\begin{equation}
\nabla_y^2 A =
\left\{\begin{array}{l} A^{\prime\prime}, \qquad ~~~~~~~~~~~~~ (\epsilon=0)\\
A^{\prime\prime}+5 A^\prime \cot{y}, \quad ~(\epsilon=+1)\\
A^{\prime\prime}+5 A^\prime \coth{y},\quad (\epsilon=-1)
\end{array} \right.
\end{equation}
Let us first consider a specific example, with $\alpha=0$ and
$\beta=2$. For simplicity, we set $\tau=\rho=1$.
Equations~(\ref{symm-4D}) and (\ref{symm-6D}) reduce
to~\footnote{Even though we specified above the metric of the
compact 6D space, the results here are valid for any other 6D
spaces, including Ka\"hler manifolds, with an arbitrary curvature.
Einstein's equations carry no information about the topology of
internal manifold but only its spatial curvature.}
\begin{equation}\label{10dmunu}
{}^{(10)}R_\mu^\mu = e^{-2A} \hat{R}_4 - 16(\nabla_y{A})^2 - {4}
\nabla_y^2{A},
\end{equation}
and
\begin{eqnarray}\label{10dij}
{}^{(10)}R_{i}^{i} &=& \tilde{R}_{6} - 4\nabla_i
\partial^i {A} -4 \nabla_i A \partial^i A.
\end{eqnarray}
Note that, since $A\equiv A(y)$, the last two terms above
contribute only when $i=y$.

Now, from the trace-subtracted 10D Einstein equations, i.e.
\begin{equation}
{}^{(10)} R_{AB} = T_{AB}-\frac{1}{8} g_{AB} T_C^C,
\end{equation}
we obtain
\begin{equation}\label{10dmunu-2}
{}^{(10)}R_\mu^\mu = \frac{1}{2} T_\mu^\mu-\frac{1}{2} T_m^m
\equiv \frac{1}{2} T_4 -\frac{1}{2} T_6,
\end{equation}
\begin{equation}\label{10dij-2}
{}^{(10)}R_i^i = \frac{1}{4} T_i^i-\frac{3}{4} T_\mu^\mu \equiv
\frac{1}{4} T_6 -\frac{3}{4} T_4.
\end{equation}
To return to the familiar 4D notations, one may replace $T_{AB}$
by $8\pi G {\cal T}_{AB}$. Here we also introduce the following
notations:
\begin{equation}
\hat{R}_4 = {}^{(4)}\hat{R}_\mu^\mu, \qquad \tilde{R}_6 =
{}^{(6)}\tilde{R}_i^i, \qquad R^{(6)}\equiv {}^{(10)} R_i^i.
\end{equation}
In a sense, $R^{(6)}$ is the total integrated 6D scalar curvature,
which is generally positive, while $\tilde{R}_6$ is the scalar
curvature associated with the 6D metric itself.

Equating Eqs.~(\ref{10dmunu}) and (\ref{10dij}) with
Eqs.~(\ref{10dmunu-2}) and (\ref{10dij-2}), we find
\begin{equation}
\tilde{R}_6 = - 12 (\nabla_y{A})^2 + e^{-2A} \hat{R}_4 +
\frac{3}{4} {T}_6 - \frac{5}{4} {T}_4.
\end{equation}
Similarly, with $\beta=-\alpha=2$, we find
\begin{equation}
\tilde{R}_6 = 8 (\nabla_y{A})^2- 10 \nabla_y^2{A} + e^{-4A}
\hat{R}_4 + e^{-2A} \left(\frac{3}{4} {T}_6 - \frac{5}{4}
{T}_4\right).
\end{equation}
As is evident, different choices of $\beta$ and $\alpha$ can
easily lead to different set of equations.

Of course, the above set of equations do not lead to a 4D de
Sitter solution unless that the energy momentum tensor $T_{AB}$
has components that violate certain (positivity) conditions. In a
general $D$ dimensions, with $m$ extra dimensions, i.e. $D=d+m$,
the strong energy condition is $T_m\ge \frac{m-2}{d}
T_d$~\cite{Gibbons-84,Ish01c}, which with $d=4$ and $m=6$ reads
$T_6\ge T_4$~\footnote{In~\cite{DK10}, Douglas and Kallosh
introduced a new condition, so-called internal energy condition,
$T_m\ge \frac{m}{d-2} T_d$, which can however be violated in
$D=10$ by wrapped Dp-branes when $p>7$. This is also a reason for
why one considers, for instance, in type IIB string theory, only
Dp-branes with $p\le 7$.}.

We have not specified yet the source of energy momentum tensor
fields, which may include contributions from a bulk cosmological
constant, fluxes or even some non-local effects of warped branes
in the extra dimensions, but their explicit expressions won't be
important for our discussion in this section. We just want to
emphasize here that the condition like $T_6\ge T_4$ (in $D=10$
dimensions) does not necessarily enforce $\tilde{R}_6$ only to
take a positive value; the other two choices ($\tilde{R}_6=0$ and
$\tilde{R}_6<0$) are also possible.

\subsection{A more general choice of 6D metric}

For the choice of 6D metrics as in
(\ref{6d-flat})--(\ref{6d-negative}), there is no free parameter
which may be tuned to the coefficient such as $\alpha$ in the
volume factor, which  could otherwise lead to a de Sitter solution
in four dimensions. If we write the 6D metric ansatz in the
following form:
\begin{equation}
ds_6^2 = g(y) dy^2 + \alpha_1 f(y)
d\Omega_5^2,\label{6d-two-terms}
\end{equation}
then we immediately see that a four-dimensional de Sitter solution
can be obtained by tuning the coefficient $\alpha$ with the
constant $\alpha\Z{1}$ related to the 6D curvature. This is
possible also for a canonical choice that $g(y)=1$ and $f(y)=y^2$.
A detailed analysis actually shows that if we set $\alpha=0$ (from
the beginning) and allows the 6D metric to take the form
of~(\ref{6d-two-terms}), then the solution for the warp factor
must take a rather nontrivial form in order to yield de Sitter
solutions in four-dimensions. From this we also conclude that even
though one could in principle absorb the volume factor $e^{\alpha
A(y)}$ within the internal metric (cf equation~\ref{main-10d-anz})
or set $\alpha=0$, there are no advantages with this choice.

\subsection{Nontrivial solutions}

In the analysis below, we write the 6D metric ansatz in the
following form:
\begin{equation}\label{main-6d-gen}
ds_6^2= \sinh^2{y}\, dy^2 + \alpha_1 \cosh^2{y}\, ds_{X_5}^2.
\end{equation}
where $X_5$ is either $S^5$ or an Einstein--Sasaki space
$T^{1,1}=(S^2 \times S^2)\rtimes S^1$. As compared to the metrics
(\ref{6d-flat})--(\ref{6d-negative}), we now have one more free
parameter, $\alpha_1$. The metric~(\ref{main-6d-gen}) is Ricci
flat only if $\alpha_1=1$, while it is positively (negatively)
curved for $\alpha_1<1$ ($\alpha_1>1$). There are two motivations
behind the above choice of the metric. First, unlike in the
examples considered in~\cite{KT}, the size of the compact space
$X_5$ is nonzero (avoiding a conical singularity at $y=0$) and the
curvature components are regular everywhere. Second, the problem
of finding four-dimensional de Sitter solutions is well posed,
especially, if all extra dimensions are physically compact, so we
can study higher-dimensional supergravity theories by allowing a
noncompact dimension. This approach is perfectly viable and is
also consistent with the observation that the Calabi--Yau spaces
in string theory are generically noncompact and they are also
known to allow at least one noncompact direction, admitting a
warped throat geometry.

Of course, using the coordinate transformation $\sinh{y}\, dy \to
dz $, we may express (\ref{main-6d-gen}) into a more familiar form
$ ds_6^2= dz^2 + f(z)^2\, ds_{X_5}^2$, where $f(z) \propto (z+
c)$, but we prefer to use the metric (\ref{main-6d-gen}) as it is
sufficiently general and also convenient for the purpose of
solving Einstein's field equations. With (\ref{main-6d-gen}), we
find that the Ricci tensor components of the 10D spacetime are
related to those in 4D spacetime and the internal spaces by
\begin{subequations}
\begin{align}
{}^{(10)} R_{\mu\nu}(x,y) = {}^{(4)}\hat{R}_{\mu\nu}(x)-
\frac{\hat{g}_{\mu\nu}\,e^{(\beta-\alpha)A}}{\sinh^2{y}}
\left[(\beta^2+\beta\alpha){A^\prime}^2+\frac{\beta}{2}
A^{\prime\prime}+\frac{\beta}{2}\left(5\tanh{y}-\coth{y}\right)A^\prime\right]
\label{eqn-R1},\\
R_{yy} = - \frac{4\beta+5\alpha}{2}
A^{\prime\prime}-(\beta^2-\beta\alpha){A^\prime}^2+2\beta A^\prime
\coth{y}+
\frac{5\alpha A^\prime}{\sinh(2y)},\label{eqn-R2}\\
{}^{(10)} R_{p q} = \tilde{R}_{p q} - \tilde{g}_{p q}\,
\alpha_1\coth^2{y} \left(\frac{\alpha
A^{\prime\prime}}{2}+(\alpha\beta+\alpha^2) {A^\prime}^2+
(2\beta+4\alpha)A^\prime \tanh{y} -\frac{\alpha
A^\prime}{\sinh{2y}}\right),\label{eqn-R3}
\end{align}
\end{subequations}
where ${}^\prime \equiv \partial/\partial {y}$ and
\begin{equation}
{\tilde R}_{p q} \equiv 4\left(1 - \alpha_1\right) \tilde{g}_{pq}.
\end{equation}
Here $\tilde{g}_{pq}$ denote the metric components of the base
space $X_5$, such as $S^5$ or $T^{1,1}$, which are independent of
the y coordinate. In the $\alpha\ne \beta$ case, we shall
consistently choose
\begin{equation}
\alpha_1= \frac{(\beta-\alpha)^2}{2\beta^2}.
\end{equation}
At this stage, we make no {\it prior} assumptions about the
internal space curvature. 

\subsection{Positive curvature}

Take $\beta=2$, $\alpha=4$ and $\alpha_1=1/2$. The internal 6D
space with the metric (\ref{main-6d-gen}) is positively curved,
$\tilde{R}_{pq} = 2 \tilde{g}_{pq}$. The components of the 10D
Ricci curvature are given by
\begin{subequations}
\begin{align}
{}^{(10)} R_\mu^\mu= {}^{(4)} \hat{R}_\mu^\mu \, e^{-2A} -
\frac{4\,e^{-4A}}{\sinh^2{y}}\left( 12{A^\prime}^2+ \nabla_y^2 A
\right) = \frac{1}{2} \left(
{T}_4-{T}_6\right),\label{a0b2-a} \\
{}^{(10)} R_{i}^i= {}^{(6)} \tilde{R}_i^i \,e^{-4A}-
\frac{4\,e^{-4A}}{\sinh^2{y}}\left( 29 {A^\prime}^2 + 6 \nabla_y^2
A \right) = \frac{1}{4} \left( {T}_6-3{T}_4\right),\label{a0b2-b}
\end{align}
\end{subequations}
where
\begin{equation}
\tilde{R}_6 = {}^{(6)}\tilde{R}_i^i=
\frac{20(1-\alpha_1)}{\alpha_1\,\cosh^2{y}}
\end{equation}
and
\begin{equation} \nabla_y^2 A =
A^{\prime\prime} + \left(5\tanh{y}-\coth{y}\right)A^\prime.
\end{equation}
From equations~(\ref{a0b2-a}) and (\ref{a0b2-b}), one has
\begin{equation}
\tilde{R}_6 = \frac{4}{\sinh^2{y}}\left(17\,{A^\prime}^2 + 5
\nabla_y^2 A\right) + e^{2A} \hat{R}_4 + \left(\frac{3}{4} {T}_6 -
\frac{5}{4} {T}_4\right) e^{4A}.
\end{equation}
Unlike with the maximally symmetric metrics introduced in
(\ref{6d-flat})-(\ref{6d-negative}), which are singular at $y=0$,
for the metric~(\ref{main-6d-gen}), the 6D curvature tensors can
be regular everywhere. 

Here, we take $T_{AB}=0$, so that the condition such as $T_6\ge
T_4$ is trivially satisfied (also saturated), $T_6=T_4=0$. The 10D
Einstein equations are then explicitly solved when
\begin{equation}
A(y) = - \ln \cosh{y} - A_0, \quad \hat{R}_4 = 32\,e^{2A_0}, \quad
\tilde{R}_6 = \frac{20}{\cosh^2{y}}.
\end{equation}
Note that the warp factor $e^{A(y)}\to \infty$ when $A_0\to
-\infty$, which implies that a flat 4D Minkowski spacetime is not
a solution in the above case. The requirement that the warp factor
is real and positive definite also rules out an anti-de Sitter
solution. Further, if the 4D metric takes the form of a standard
FRW universe
\begin{equation}\label{FRW}
ds_4^2  \equiv \hat{g}_{\mu\nu} dx^\mu dx^\nu =  -dt^2+
a^2(t)\left[\frac{dr^2}{1-k r^2}+ r^2 d\Omega_2^2\right],
\end{equation}
where $k$ is the 3D spatial curvature, then the scale factor is
given by
\begin{equation}
a(t)= \frac{a_0}{2} \exp\left(\sqrt{\frac{8}{3}}\,e^{A_0}\,
t\right) + \frac{3k}{16 a_0} \,e^{-2A_0}
\exp\left(-\sqrt{\frac{8}{3}}\,e^{A_0}\, t\right),
\end{equation}
where $a_0$ is an arbitrary constant. This is nothing but a de
Sitter solution in an accelerating patch~\footnote{ The universe
we live in was perhaps not de Sitter at all times. It is easy to
understand why this might be the case because the presence of
matter fields at later epochs of cosmic evolution can easily lead
the expansion away from a pure de Sitter phase. As we discuss
below, the consideration of standard background fluxes, or p-form
gauge fields, which contribute to $R^{(4)}$ negatively, can also
do this job.}. Note that, unlike in pure supergravity theories
with a flat 3-brane, for which a nontrivial warp factor only
arises as the back-reaction of some fluxes (and hence a
non-vanishing $T_{AB}$), the vacuum energy on a de Sitter brane
(or an inflating FRW universe) can naturally warp the bulk
spacetime and introduce a nontrivial warp factor even if
$T_{AB}=0$.

\subsection{Zero curvature}

Next, we make the choice $\alpha=(1+ \sqrt{2})\beta$ and
$\alpha_1=1$. The internal 6D space with the metric
(\ref{main-6d-gen}) is Ricci flat, $\tilde{R}_{pq} = 0$. We then
find
\begin{subequations}
\begin{align}
\hat{R}_4\, e^{-\beta A}  - \frac{e^{-(1+ \sqrt{2})\beta
A}}{\sinh^2{y}}\left((8+ 4\sqrt{2})\beta^2 (\nabla A)^2+2\beta
\nabla^2 A\right) = \frac{1}{2} \left(
{T}_4-{T}_6\right), \\
e^{-(1+ \sqrt{2})\beta A} \left( \tilde{R}_6 -
\frac{1}{\sinh^2{y}} \left( \left(20+ 14\sqrt{2}\right)\beta^2
(\nabla A)^2 + (7+ 5\sqrt{2})\beta\, \nabla^2 A\right)\right) =
\frac{1}{4} \left( {T}_6-3{T}_4\right).
\end{align}
\end{subequations}
From these equations we derive
\begin{eqnarray}
\tilde{R}_6 &=&  \frac{1}{\sinh^2{y}}\, \left( \beta^2 \left(12+
10\sqrt{2}\right) (\nabla A)^2+ 5(1 + \sqrt{2}) \nabla^2
A\right)+ e^{ \sqrt{2}\,\beta A} \hat{R}_4 \nonumber \\
&{}& \qquad \, + \left(\frac{3}{4} {T}_6 - \frac{5}{4}
{T}_4\right) e^{(1+ \sqrt{2})\beta A}.\label{6-and-4-curvs-2}
\end{eqnarray}
With $T_{AB} =0$, the 10D Einstein equations are explicitly solved
for
\begin{eqnarray}
&& a(t)= \frac{a_0}{2} \exp\left(\sqrt{\frac{4}{3}}\,e^{A_0}\,
t\right) + \frac{3k}{8 a_0} \,e^{-2A_0}
\exp\left(-\sqrt{\frac{4}{3}}\,e^{A_0}\, t\right), \nonumber\\
&& A(y) =- \frac{1}{\sqrt{2}} \left(\ln \cosh{y}
+A_0\right),\qquad \tilde{R}_6 = 0.
\end{eqnarray}

\subsection{Negative curvature}

Finally, we take $\beta=2$, $\alpha =6$ and $\alpha_1=2$. The
internal 6D space with the metric (\ref{main-6d-gen}) is
negatively curved, $\tilde{R}_{pq} = -4 \tilde{g}_{pq}$. We then
find
\begin{subequations}
\begin{align}
{}^{(10)} R_\mu^\mu= {}^{(4)} \hat{R}_\mu^\mu\, e^{-2A}  -
\frac{4\,e^{-6A}}{\sinh^2{y}}\left(16 (\nabla_y A)^2 + 4
\nabla_y^2 A\right) = \frac{1}{2} \left(
{T}_4-{T}_6\right), \\
{}^{(10)} R_{i}^i= {}^{(6)} \tilde{R}_i^i \, e^{-6A}-
\frac{4\,e^{-6A}}{\sinh^2{y}} \left( 58 (\nabla_y A)^2 +
\frac{17}{2} \nabla_y^2 A\right) = \frac{1}{4} \left(
{T}_6-3{T}_4\right).
\end{align}
\end{subequations}
From these equations we derive
\begin{equation}
\tilde{R}_6 =  \frac{4}{\sinh^2{y}}\, \left( 42 (\nabla_y A)^2+
\frac{9}{2} \nabla_y^2 A\right) + e^{4A} \hat{R}_4 +
\left(\frac{3}{4} {T}_6 - \frac{5}{4} {T}_4\right)
e^{6A}.\label{6-and-4-curvs}
\end{equation}
With ${T}_{AB}=0$, the 10D Einstein equations are explicitly
solved when
\begin{eqnarray}
&& a(t)= \frac{a_0}{2} \exp\left(\sqrt{\frac{2}{3}}\,e^{A_0}\,
t\right) + \frac{3k}{4 a_0} \,e^{-2A_0}
\exp\left(-\sqrt{\frac{2}{3}}\,e^{A_0}\, t\right), \nonumber \\
&& A(y) =- \frac{1}{2} \left( \ln \cosh{y} + A_0\right), \quad
\tilde{R}_6 = -\frac{10}{\cosh^2{y}}.
\end{eqnarray}

Coming to the issue of the Maldacena-Nunez no-go theorem, the main
assumption in~\cite{Malda-Nunez} appeared to be that the internal
manifold is physically compact and the integrand $\int \nabla^2
e^{n A} $ vanishes, where $n$ is some constant. This last
condition can be simply relaxed if the extra-dimensional manifold
is only geometrically compact or when there exists a noncompact
direction (as above). In fact, the quantities such as $\nabla^2 A$
and $\nabla^2 e^{nA(y)}$ encode information about the spatial
curvature of the internal metric. As a consequence, depending upon
the choice of the 6D curvature, the integrand $\int \nabla^2
e^{nA}$ may or may not vanish. The other crucial assumption of the
MN no-go theorem was that the extra-dimensional warped volume is
finite (or a finite Newton's constant), which is however not
violated by the solutions presented in this paper.

\subsection{Combining results}

The above results can be expressed in a more illustrative and
simpler form:
\begin{equation}
ds_{10}^2 = \frac{e^{2{\cal A}}}{\left(\cosh{y} + \epsilon
\sinh{y}\right)^{2\lambda}}\left[ ds_4^2 + \frac{4 L^2}{3}
\Big(2\lambda^2\left(\frac{\epsilon+\tanh{y}}{1+\epsilon\tanh{y}}\right)^2
dy^2 + d\Omega_{X_5}^2\Big)\right], \label{3cases-com}
\end{equation}
where $\lambda$ and ${\cal A}$ are dimensionless constants,  $L$
has the dimension of length and
\begin{equation}
ds_4^2 =  -dt^2+ a^2(t)\left[\frac{dr^2}{1-k r^2}+ r^2
d\Omega_2^2\right],\quad a(t)= \frac{a_0^2+k L^2}{2 a_0}
\cosh\frac{t}{L} + \frac{a_0^2-kL^2}{2a_0} \sinh\frac{t}{L},
\end{equation}
where $a_0$, as above, is an arbitrary constant. The metric
(\ref{3cases-com}) is an exact solution to 10D Einstein equations
without sources ($T_{AB}=0$). The solutions presented in the above
subsections are obtained by setting $\epsilon=0$, taking a
specific value of $\lambda$ and then suitably redefining some of
the numerical constants. The above results are new and they are
quite remarkable. For the reasons to be explained below we chose $
0\le \epsilon^2<1$ and $\lambda>0$.

As briefly mentioned above, the de Sitter solutions presented here
get around the no-go theorem because at least one of the
assumptions of the original no-go theorem is violated. Indeed, we
have chosen to `compactify' a string-inspired supergravity model
by considering a non-compact dimension. This may sound
counterintuitive, but actually it is a well-defined procedure
known as consistent warped compactification. To be precise, we
will explicitly show that the extra-dimensional warped volume is
finite for our solutions and hence the four-dimensional effective
Planck mass is also finite.

On passing, we note that the 6D volume (without the effect of
warping) is
\begin{equation}
V_6 = \int \sqrt{g_6}\, d^6 x = \frac{64\sqrt{2}\,L^6}{27} ~{\rm
Vol}(X_5)\int \frac{\epsilon +\tanh{y}}{1+\epsilon \tanh{y}}\, dy,
\end{equation}
where ${\rm Vol} (X_5)=16\pi^3/27$ if $X_5=T^{1,1}$ and ${\rm Vol}
(X_5)=8\pi^2/15$ if $X_5=S^5$. This integral can be explicitly
evaluated. We find that
\begin{equation}
V_6 \propto \left(\ln \cosh{y}+ \ln(1+\epsilon \tanh{y})\right).
\end{equation}
We can clearly see the necessity of the choice $\epsilon <1$,
especially, in the case y is allowed to range from $-\infty$ to
$+\infty$ so that the 6d volume is regular everywhere. From the
solution (\ref{3cases-com}) itself it is clear that, in the large
volume limit $y\to \infty$, the radius modulus, which scales as
$|(\epsilon +\tanh{y})/(1+\epsilon \tanh{y})|$, takes a finite
value. Of course, the volume modulus $\sigma \equiv \left({\rm
Vol}_6\right)^{1/3}$ grows linearly with $y^{1/3}$, since
$|\tanh{y}|\to 1$ as $y\to \infty$. This growth could possibly be
halted or minimized, leading to a rough stabilisation of extra
dimensional volume, by adding additional energy sources, such as
wrapped branes and fluxes, or even some non-perturbative
contributions to the vacuum expectation value of $\langle
\sigma\rangle $ as in Kachru-Kallosh-Linde-Trevedi (KKLT)-type
constructions~\footnote{In such a case, the solution to 10D
supergravity equations may not be exactly of de Sitter type, since
additional sources or fluxes can modify both the warp/conformal
factors and the 4D curvature.}; we may not even require any such
corrections for the purpose of getting a finite warped volume or a
finite four-dimensional Newton's constant.

\begin{figure}[!ht]
\centerline{\includegraphics[width=2.7in,height=2.0in]
{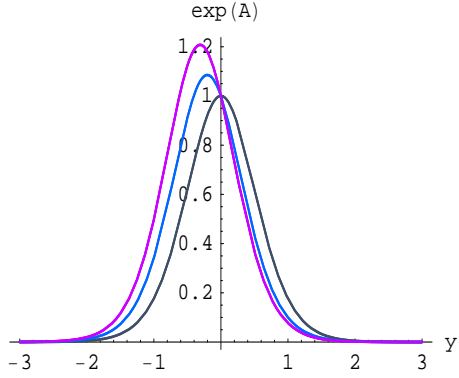}} \caption{The plot of the warp factor
$\exp(A(y))=\exp({\cal A})/(\cosh{y}+\epsilon \sinh{y})^\lambda$
with ${\cal A}=0$, $\lambda=2$ and $\epsilon=0.3, 0.2, 0$ (from
top to bottom) (online: pink, blue and black).} \label{3dvolume}
\end{figure}

Note that the warp factor
$$ e^{A(y)} \equiv \frac{e^{{\cal A}}}{(\cosh{y}+
\epsilon \sinh{y})^{\lambda}}$$ is bounded from the above and
below when $\lambda>0$ and $0\le \epsilon <1$ (cf
figure~(\ref{3dvolume})). Around $y\simeq 0$, we have
\begin{equation}
e^{2A} \sim \exp\left( -2\lambda |y|\right).
\end{equation}
As is evident, the larger is the value of $\lambda$ (and hence the
longer the throat region), the stronger would be the warping along
the transverse direction. For a given $\lambda$ (which we take to
be positive) the warped volume would be minimum when $\epsilon=
0$. This is more evident from the following explicit result:
\begin{eqnarray}
\int d^{10} x \sqrt{-g_{10}} R_{10} & = & \frac{64\sqrt{2}L^6
\lambda}{27}\,e^{8{\cal A}}  \int d\Omega_5 \int_{y\Z0}^{y\Z1}
\frac{\sinh{y}+\epsilon \cosh{y}}{(\cosh{y}+\epsilon
\sinh{y})^{8\lambda+1}}\, dy
\int d^4{x} \sqrt{\hat{g}} \left( \hat{R}_4-\frac{12}{L^2} \right)\nonumber \\
&= & \frac{8\sqrt{2} L^6}{27}\,e^{8{\cal A}} {\rm Vol} (X_5) \,
\left[\frac{-1}{(\cosh{y}+\epsilon
\sinh{y})^{8\lambda}}\right]_{y\Z0}^{y\Z1} \int d^4{x}
\sqrt{\hat{g}_4} \left( \hat{R}_4- \frac{12}{L^2}
\right).\nonumber\\
\end{eqnarray}
As long as $\epsilon^2 <1$ and $\lambda>0$, the 6D warped volume
is finite. A large and positive $\lambda$ implies a strong warping
of extra spaces and also leads to a relatively large throat region
(internal volume)~\footnote{In~\cite{Greene10}, Greene et al.
argued that models of compactifications with both a large volume
and a large mass gap may produce phenomenologically acceptable
inflationary models with less number of fine tunings, and this
might actually be the case also in the present construction once
one supplements the gravitational action with a scalar
potential.}. Let us take $\epsilon=0$, just for simplicity. We
then obtain
\begin{equation}
\frac{M_{(10)}^8}{(2\pi)^6}\int d^{10}{x}\, \sqrt{-g_{10}}\,R_{10}
=M_{\rm Pl}^2 \int d^4{x}\, \sqrt{-g_4} \left(R_4
-\Lambda_4\right),
\end{equation}
where $M_{(10)}$ is the fundamental 10D Planck scale, $\Lambda_4
\equiv 12/L^2$ and
\begin{equation}
M_{\rm Pl}^2= \frac{M_{(10)}^8\, e^{8{\cal A}}}{(2\pi)^6}\,
V_6^{\rm w}, \qquad V_6^{\rm w}\equiv {\rm Vol}
(X_5)\,\frac{8\sqrt{2} L^6}{27} \times
\frac{2}{\cosh(y)^{8\lambda}}\Big{\vert}_{y\Z0}^{y\Z1}.\label{def-Newton}
\end{equation}
Note that one does not have to introduce any artificial cutoff to
make the Newton's constant finite in (\ref{def-Newton}); the
integrand is finite even when we take $y\Z0=-\infty$ and
$y\Z1=+\infty$. Of course, one may take $y_0= y_{\rm IR}$ and
$y\Z1= y_{\rm UV}$, without loos of generality. Moreover, the size
of the radial (fifth) dimension, along with the size of compact
$X_5$ manifold, takes a constant value in the large volume limit,
providing an explicit example of spontaneous compactification. The
stability of the compactified space is guaranteed at least at a
classical level. We have also checked that there is a zero-mode
graviton wavefunction that is normalizable in the usual $3+1$
spacetime~\cite{Ish10e}. In a follow up work we will study in
detail the effects of continuum KK modes on a 4D gravitational
potential and related physics, where we also take into account the
effects of a scalar field Lagrangian.

It may be worth pointing out a few crucial differences in the
results between an earlier paper by Gibbons and Hull
(GH)~\cite{GH2001} and the one here: GH failed to find, within a
class of non-compact warped supergravity solutions, a sensible 4D
de Sitter solution that leads to a finite 4D Planck mass, whereas
the one here has achieved this goal. Moreover, within the GH
construction, the viable solutions were found to be singular,
which were realised by considering a geometry for which the
internal space is asymptotic to a cone over a product of spheres,
whereas the solutions presented here are non-singular.

\subsection{Neglecting the warp factor}

In this subsection, we momentarily return to the set of 10D
curvature tensors presented in section 2, i.e. equations
(\ref{symm-4D}) and (\ref{symm-6D}), from which we obtain
\begin{subequations}
\begin{align}
{}^{(10)}R_\mu^\mu & = e^{-\beta A} \hat{R}_4 - e^{-\alpha A}
\left(4(\beta^2+\beta\alpha) (\nabla A)^2 + 2\beta \nabla^2
A\right) = \frac{1}{2} T_4 - \frac{1}{2} T_6,\\
{}^{(10)}R_{i}^{i} & = e^{-\alpha A} \left(\tilde{R}^{(6)}-
(5\alpha^2 +4\beta\alpha +\beta^2)(\nabla
A)^2-(2\beta+5\alpha)\nabla^2 A\right) = \frac{1}{4} {T}_6 -
\frac{3}{4} {T}_4.
\end{align}
\end{subequations}
From these equations, along with equation~(\ref{symm-4D}), we
derive
\begin{eqnarray}
\hat{G}_{\mu\nu} &=&\hat{R}_{\mu\nu}-\frac{1}{2}\hat{g}_{\mu\nu}
\hat{R}_4 \nonumber \\
&= & {T}_{\mu\nu} + \frac{\hat{g}_{\mu\nu}}{2} \left(\frac{1}{4}
{T}_6 - \frac{3}{4} {T}_4\right) e^{\beta A} - \hat{g}_{\mu\nu}
\left((\beta^2+\beta\alpha) (\nabla A)^2 +\frac{\beta}{2}
\nabla^2{A}\right) e^{(\beta-\alpha)A}  \nonumber \\
&=& {T}_{\mu\nu} + \frac{\hat{g}_{\mu\nu}}{2} \left(
\tilde{R}^{(6)}- \left( 5\alpha^2+6\beta\alpha+ 3\beta^2 \right)
(\nabla A)^2 - (3\beta+5\alpha) \nabla^2{A} \right)
e^{(\beta-\alpha)A}.\label{4D-Eins-fin}
\end{eqnarray}
In the limit $A(y)\to 0$, or that $A(y) \to A_0$, in that later
case the constant term $e^{(\beta-\alpha)A_0}$ can be absorbed
into $\tilde{R}^{(6)}$), the above expression reduces to (see
also~\cite{DK10})
\begin{equation}
\hat{R}_{\mu\nu} - \frac{1}{2} \hat{g}_{\mu\nu} \hat{R} =
{T}_{\mu\nu} + \frac{1}{2} \hat{g}_{\mu\nu}
\tilde{R}^{(6)}.\label{munu-4D}
\end{equation}
Evidently, in the limit $A(y)\to 0$, one has $R_4 = \hat{R}_4$,
$R_6= \tilde{R}_6$ and hence
\begin{equation}
\hat{R}_4 = \frac{1}{2} \left({T}_4 - {T}_6\right), \qquad
\tilde{R}_6 = \frac{1}{4} \left({T}_6 - 3{T}_4\right),
\end{equation}
and
\begin{equation}
\hat{R}_4 = -{T}_4 - 2 \tilde{R}_6.\label{rel-4D-6D}
\end{equation}
This result shows that in a universe dominated by radiation, for
which $a(t)\propto t^{1/2}$ and ${T}_4 =0$, the magnitude of the
4D Ricci scalar is just two times the magnitude of 6D Ricci
scalar, but with opposite sign. This implies that, in the absence
of warping, the extra dimensional manifold remains equally large
as the physical 4D universe, leading to a phenomenologically
unacceptable scenario.

A pertinent question to ask is: Can we get de Sitter solutions
just by adding background fluxes, but dropping the warp factor?
The answer seems to be negative despite the occurrence of one or
more extra terms on the right-hand side of
equation~(\ref{munu-4D}) or equation~(\ref{rel-4D-6D}).

\section{Effects of flux}

In this section, we consider a $D$-dimensional Einstein action
coupled to p-form gauge field strengths or matter fields:
\begin{equation}
S=  \int \sqrt{-g} \left(R^{(D)} - \frac{1}{2} \sum_{p}
|F_{(p)}|^2 \right).
\end{equation}
Here, as in~\cite{WSH87,DK10}, we follow a non-standard
normalisation of $|F_{(p)}|^2$, which allows one to treat the
$p=0$ and $p>0$ cases uniformly.

\subsection{Magnetic flux}

The energy--momentum tensor due to p-form field strengths is given
by
\begin{equation}
T_{AB}= p F_{A Q_1 Q_2 \cdots Q_{p-1}} F_{B}\,^{Q_1 Q_2 \cdots
Q_{p-1}} - \frac{1}{2} g_{AB} F_p^2.
\end{equation}
With p-form magnetic flux (M), only the ($ij$) components of
$F_{p}$ are non-vanishing. Since $F_p$ depends only on internal
space coordinates, we obtain
\begin{eqnarray}
T_{\mu\nu}^{(M)}= -\frac{1}{2} g_{\mu\nu} F_p^2, \quad T_{ij} = p
F_{i m_1 m_2 \cdots m_{p-1}} F_{j}\,^{m_1 m_2 \cdots
m_{p-1}}-\frac{1}{2} g_{ij} F_p^2.
\end{eqnarray}
This implies that $T_4^{(M)}= - 2 F_p^2$, $T_6^{(M)} = (p-3)
F_p^2$ and hence
\begin{equation}
R_6=\frac{1}{4} T_6 -\frac{3}{4} T_4 =
\frac{p+3}{4}\,F_p^{2}.\label{R6-mag}
\end{equation}
That is, a magnetic flux contributes to $R_6$ positively.

\subsection{Electric flux}

Here we consider the effect of an electric flux. In the case of
p-form electric flux (E) with $p\ge 4$, an appropriate ansatz is
\begin{equation}
{F}_{\mu\lambda\rho\sigma q_1 q_2 \cdots q_{p-4}}=i
\epsilon_{\mu\lambda\rho\sigma} \,{f}_{q_1 q_2 \cdots q_{p-4}}.
\end{equation} The four components (or legs) of
$F_p$ are in usual 4D spacetime and the rest are in internal
spaces. This yields
\begin{subequations}
\begin{align}
T_{\mu\nu}^{(E)}= -\frac{1}{2} g_{\mu\nu} p (p-1) (p-2) (p-3)
f_{q_1\cdots q_{p-4}}f^{q_1\cdots q_{p-4}} \equiv -\frac{1}{2}
g_{\mu\nu} \tilde{f}_p^{\,2},\\
T_{ij}^{(E)}= -p(p-1)(p-2)(p-3) (p-4) f_{i q_1 q_2 \cdots q_{p-5}}
f_{j}\,^{q_1 q_2 \cdots q_{p-5}}+ \frac{1}{2} g_{ij}
\tilde{f}_p^{\,2}.
\end{align}
\end{subequations}
From these we obtain
\begin{equation}
T_4^{(E)}=-2 \tilde{f}_p^{\,2}, \quad T_6^{(E)}=(7-p)
\tilde{f}_p^{\,2}
\end{equation}
and hence
\begin{equation}
R_6=\frac{1}{4} T_6 -\frac{3}{4} T_4 =
\frac{13-p}{4}\,\tilde{f}_p^{\,2} \equiv
\frac{\tilde{p}+3}{4}\,\tilde{f}_{10-\tilde{p}}^{\,2}.
\label{R6-elec}
\end{equation}
In the last line above we replaced $p$ by $(10-\tilde{p})$.
Comparing this with (\ref{R6-mag}) reveals that an electric
contribution is dual to a magnetic flux. As is evident, the
replacement
$$ p \to (10-{p}),\quad F_p^2 \to \tilde{f}_p^2 $$
is nothing but a symmetry between magnetic and electric ${p}$-form
field strengths.

In addition to the background p-form field strengths, one could
also introduce some localized objects or brane sources (with $p\ge
3$), in some subspaces of the internal manifold ${\cal M}$:
\begin{equation}
S_{\rm loc} = - \tau_p \int_{M_4\times {\cal M}} d^{p+1} \xi
\sqrt{-\tilde{g}} + \mu_p \int_{M_4\times {\cal M}} C_{p+1},
\end{equation}
where $\tau_p$ is the p-brane tension, $C_{p+1}$ represents
Chern--Simon terms and the pull-back metric $\tilde{g}$ is defined
through
$$\tilde{g}_{\mu\nu}\equiv \frac{\partial
X^A}{\partial\xi_\mu} \frac{\partial X^B}{\partial\xi_\nu}
g_{AB}.$$ In this case, $R_4$ and $R_6$ each term receives an
extra contribution:
\begin{equation}
R_{4}^{\rm loc}= \frac{p-7}{4} \tau_p\, \delta({\cal M}), \qquad
R_{6}^{\rm loc}= \frac{15-p}{4} \tau_p\, \delta({\cal M}).
\end{equation}
The existence of delta function $\delta({\cal M})$ implies that
the energy momentum tensor for the local sources is localized (or
confined) within the internal manifold ${\cal M}$ or its subspace.

\subsection{Total effects}

Combining the above results, in spacetime dimensions $D=10$, one
finds~\cite{WSH87,DK10}
\begin{subequations}
\begin{align}
{}^{(10)}R_\mu^\mu \equiv {R}_4 = -\frac{p-1}{2} F_p^2 +
\frac{p-9}{2} \tilde{f}_p^2 + \frac{p-7}{4} \tau_p \delta({\cal
M}),\label{total-eff-a}\\
{}^{(10)}R_i^i \equiv {R}_6 = \frac{p+3}{4} F_p^2 + \frac{13-p}{4}
\tilde{f}_p^2 + \frac{15-p}{4} \tau_p \delta({\cal
M}).\label{total-eff-b}
\end{align}
\end{subequations}
Here $0<p \le 6$, $4\le p \le 9$ and $p\ge 3$, respectively, in
the first, second and third terms on the right-hand side. Since
$3\le p \le 9$ (in $D=10$, a D9-brane fills up whole of the
spacetime), $R^{(6)}$ is always positive, except when $\tau_p$
takes a large but negative value. However, as we explicitly show
below, the condition $R^{(6)} >0$ does not necessarily require
that $\tilde{R}_6>0$ except when the warp factor becomes a
constant.

By combining equations~(\ref{total-eff-a}) and
(\ref{total-eff-b}), we obtain
\begin{equation} R_{10} = R_4 + R_6= \frac{5-p}{4} \left(
F_p^2 - \tilde{f}_p^{\,2}\right) + 2\tau_p\, \delta ({\cal M}).
\label{sum4and6}
\end{equation}
This shows that a self-dual 5-form field contributes to $R_4$ and
$R^{(6)}$ with opposite signs and its overall contribution to 10D
Ricci curvature is zero.

\section{Incorporating the warp factor}

Let us first consider the effects of a q-form magnetic field. In
this case, using the co-ordinate transformation, $\sinh{y}\,dy\to
dz$, it would be convenient to write the 10-dimensional metric
ansatz in the following form
\begin{equation}
ds_{4+q}^2 = e^{\beta A(z)} \hat{g}_{\mu\nu} dx^\mu dx^\nu +
e^{\alpha A(z)} \left({dz}^2 + s(z)^2 dX_{q}^2\right),
\end{equation}
where $s(z)=z+{\rm const} =\cosh{y} >0 $ is a positive definite
function of $z$. Although in the string/M theory case one would
set $q=5$ (or $q=6$), let us keep it arbitrary, for generality.

From Maxwell's equation $\partial_A \left(\sqrt{-g} F^{A Q_1
\cdots Q_{q}}\right)=0$, we obtain
\begin{equation}
F^{A Q_1 \cdots Q_{q}}\propto \left(-g\right)^{-1/2}=b\, s^{-q}
\exp\left[\left(-\frac{(q+1)\alpha}{2}-2\beta \right)A\right],
\end{equation}
where $b$ is a constant. The q-form (magnetic) field strength may
be written as~\cite{Kinoshita}
\begin{equation}
F_{q}= b\, \exp\left[\left(\frac{(q+1)\alpha}{2}-2\beta\right)
A(z)\right] s^{q}(z)\, {dz}\wedge d\Omega_{q}\end{equation} and
hence \begin{equation} F_q^2 = b^2 \, e^{-4\beta A}.
\end{equation}
Through equation~(\ref{total-eff-b}), we learned that the $q$-form
fluxes (with $q\le 9$) contribute negatively to $R^{(4)}$ and
positively to $R^{(6)}$. From this one may erroneously conclude
that in the presence fluxes, there may exist a 4D de Sitter
solution only when the 6D curvature $\tilde{R}_6$ is positive.
However, as we show below, the integrated 6D curvature $R_6$ can
be positive even when the 6D space itself is negatively curved. As
a result, the positivity of flux contributions does not
necessarily kill solutions having a Ricci flat or negatively
curved 6D space.

\subsection{A specific example}

In this subsection we momentarily drop the contribution of fluxes.

The 10D metric ansatz may be written as
\begin{equation}
ds_{10}^2 = e^{\beta A}\, \hat{g}_{\mu\nu} dx^\mu dx^\nu +
e^{\alpha A} \left(e^{2B}\, dy^2 +
e^{2C}\,dX_5^2\right),\label{10d-new2}
\end{equation}
where $A, B, C$ are some functions of $y$ and
\begin{eqnarray}
dX_5^2&=& \frac{1}{9} \left(d\psi+ \sum_{i=1}^{2} \cos\theta_i
d\phi_i \right)^2 +  \frac{1}{6} \sum_{i=1}^{2}
\left(d\theta_i^2+ \sin\theta_i^2 d\phi_i^2\right),\nonumber \\
&\equiv & e_\psi^2 +  \left(e_{\theta_1}^2 + e_{\phi_1}^2+
e_{\theta_2}^2 + e_{\phi_2}^2\right).
\end{eqnarray}
Here $(\theta_1, \phi_1)$, $(\theta_2, \phi_2)$ are coordinates on
each $S^2$, $\psi$ is the coordinate of a $U(1)$ fiber, and
\begin{equation}
e_\psi= \frac{1}{3} \left(d\psi + \sum_{i=1}^2 \cos\theta_i
d\phi_i\right), \quad e_{\theta_i}=\frac{d\theta_i}{\sqrt{6}},
\quad e_{\phi_i} =\frac{\sin\theta_i d\phi_i}{\sqrt{6}}.
\end{equation}
Suppose we have $\alpha=0$. We then find that the 6D metric
becomes Ricci flat when
\begin{equation}
B(y) = C(y) + \ln \frac{d C(y)}{dy}.
\end{equation}
The two simple solutions that satisfy the above equation are given by\\
(i) $B(y)=B_0$, $C(y)=B_0+ \ln (y+c)$, \qquad
(ii) $B(y)=\ln \sinh{y}$, $C(y)=\ln \cosh{y}$.\\
For simplicity, we impose the gauge condition $\beta A\equiv
-(B+5C)$. This choice brings the 10D metric in an
Einstein--conformal frame. A straightforward calculations yields
\begin{subequations}
\begin{align}
R^{(4)} \equiv {}^{(10)} R_\mu^\mu = e^{B+5C} \hat{R}_4 - 4
e^{-2B} \left[ \frac{3}{2} {B^\prime}^2 + \frac{25}{2}
{C^\prime}^2 + 10 B^\prime C^\prime
-\frac{1}{2} (B+5C)^{\prime\prime} \right],\\
R^{(6)} \equiv {}^{(10)}R_i^i = e^{-2C} \left[ 20 + e^{2(C-B)}
\left(2 B^{\prime\prime} - 5 {C^\prime}^2 - 3 {B^\prime}^2
\right)\right],
\end{align}
\end{subequations}
where ${}^\prime \equiv \partial/\partial{y}$. As an illustrative
case, let us take~\footnote{This choice does not particularly help
us to solve the 10D supergravity equations analytically, but it is
nonetheless useful for studying some qualitative features of 4D de
Sitter solutions by allowing the 6D manifold to take a negative
($c^2>1$), zero ($c^2=1$) or positive ($c^2<1$) Ricci curvature.}
$$ B(y) \equiv \ln \sinh{y}, \qquad C(y) \equiv \ln \cosh {y} +\ln
c.$$ The 6D Ricci curvature and $({\rm Riemann})^2$ terms are now
given by
\begin{equation}
\tilde{R}_6=\frac{20(1-c^2)}{c^2\cosh^2{y}}, \quad
R_{ijkl}R^{ijkl}= \frac{8(5c^4-10c^2+17)}{c^4 \cosh^{4}y}.
\end{equation}
Furthermore,
\begin{equation}
R^{(6)}  = \frac{20}{\cosh^2{y}} \frac{1}{c^2} +
\frac{2}{\sinh^2{y}} - \frac{5}{\cosh^2{y}}- \frac{5
\cosh^2{y}}{\sinh^4{y}},
\end{equation}
and
\begin{equation}
R^{(4)}_{y\to \infty} \to \frac{\hat{R}_4}{64}\,e^{6y} -
\left(416-\frac{20}{c^2}\right) e^{-2y}, \quad R^{(6)}_{y\to
\infty} \to \left(\frac{80}{c^2} -32\right) e^{-2y}.
\end{equation}
In the range $0< c^2<5/2$, which includes all three possibilities:
$\tilde{R}_6> 0$ ($0< c^2<1$), $\tilde{R}_6= 0$ ($c^2=1$) and
$\tilde{R}_6<  0$ ($1< c^2< 5/2$), the integrated 6D curvature
$R^{(6)}$ is positive. In this particular example, $R^{(6)}$ takes
a negative value as $y\to 0$. Based on the
result~(\ref{total-eff-b}), one sees the necessity of certain
localized sources, such as, $O6$ planes or negative tension
objects. Nevertheless, in the limit $y\to 0$, since
$\sqrt{g_{6}}\to 0$, the supergravity approximations also break
down. For consistency, one then needs to supplement the
Einstein-Hilbert term in the action by higher curvature and/or
higher derivative terms with some coefficients or coupling
constants, whose positivity is not guaranteed. That is, a
gravitational Lagrangian with higher curvature terms (or stringy
corrections) will not in general lead to a strong restriction on
the sign of $R^{(6)}$.

\subsection{A more general example}

The 4D de Sitter solutions can be available in a more general case
as well. Consider the following 10D action:
\begin{equation}
S_{10} =\frac{1}{2\kappa_{10}^2}  \int \sqrt{-g_{10}} \left(R_{10}
- \frac{1}{12} F_3^2 -\frac{1}{12} H_3^2- \frac{1}{4\cdot 5!}\,
F_{5}^2 \right) + S_{\rm CS},\label{10d-sugra}
\end{equation}
where $H_3$ and $F_3$ are 3-form fields, $F_5$ is a self-dual
5-form field and $S_{\rm CS}$ denotes the type II Chern-Simons
terms. 
As compared to the
gravitational action considered in~\cite{KT}, we have set the 10D
dilaton $\Phi$ and the Ramond-Ramond scalar ${\cal C}$ to be zero.

The majority of 10D constructions made by using a conifold type
compactification of type IIB supergravity utilise a non-compact
extra dimension. However, most works along this line failed to
reproduce a sensible 4D cosmology which requires a finite 4D
Planck mass or a finite 6D warped volume. Here we report on a
class of 4D de Sitter solutions which are not only completely
regular everywhere but also lead to an explicit realisation of
four-dimensional cosmology with a finite Newton's constant.

We begin with a sufficiently general 10D metric ansatz of the form
\begin{equation}
ds_{10}^2 = e^{\beta A}\, \hat{g}_{\mu\nu} dx^\mu dx^\nu +
e^{-\beta A} \left(e^{10 B}\, dy^2 + r^2\, e^{2B}\,dX_5^2\right),
\label{10D-KT}
\end{equation}
where $r$ is an arbitrary constant, which measures the radius of
$X_5$.
A similar choice was made in the literature before, see,
e.g.~\cite{KT,Buchel-etal}. Next, using the freedom to redefine
$A(y)$, we set may $\beta=2$. The metric ansatz then takes the
following form
\begin{equation}
ds_{10}^2 = e^{2 A}\, \hat{g}_{\mu\nu} dx^\mu dx^\nu + e^{-2 A}
\left(e^{10B(y)}\, dy^2 + r^2 e^{2B(y)}
\,dX_5^2\right),\label{10d-gauged}
\end{equation}
As compared to some similar models discussed in the literature,
here we have made two important generalizations. Firstly, we have
allowed the 4D spacetime to be a maximally symmetric, implying
that it is either flat $\hat{R}_4=0$, de Sitter $\hat{R}_4 > 0$ or
anti de Sitter $\hat{R}_4 < 0$. Secondly, we have allowed an extra
(length) parameter $r$ into the 10D metric
(equation~(\ref{10d-gauged})), so that the internal 6D space is
also not necessarily flat, which is the case only if $r=1$.

The appropriate ans\"atze for the p-form fields are now given by
\begin{eqnarray}
 F_5 &=&  {\cal F} + * {\cal F},\nonumber \\
 {\cal F}&=& K(y) \left(e_{\psi} \wedge e_{\theta_1}\wedge
e_{\phi_1} \wedge e_{\theta_2} \wedge
e_{\phi_2}\right),\\
 * {\cal F}&=& \frac{e^{8A}}{r^5}\, K(y) \, \sqrt{-{\det \hat{g}_{\mu\nu}}}\,
 dy \wedge dx^0 \wedge dx^1
\wedge dx^2 \wedge dx^3,\nonumber
\end{eqnarray}
along with the following constrain equations or Bianchi
identities:
\begin{eqnarray}
&& H_3=\partial B_2 = 3 \partial_{[P}B_{QR]},\quad B_2=
f(y)\left(e_{\theta_1} \wedge e_{\phi_1}-
e_{\theta_2} \wedge e_{\phi_2}\right),\nonumber \\
&& F_3 = \partial C_2 \equiv {c_1} \,e_\psi \wedge
\left(e_{\theta_1} \wedge e_{\phi_1}- e_{\theta_2} \wedge
e_{\phi_2}\right),\quad d {}_* F_5 = dF_5= H_3\wedge F_3,
\end{eqnarray}
where $c_1$ is an arbitrary constant and f(y) a general function
of y. It is not difficult to check that, with $r^2=1$ in
equation~(\ref{10d-gauged}), the internal 6D space becomes Ricci
flat when
\begin{equation}
-B^{\prime\prime} + 2 {B^\prime}^2 + 2 e^{8B}=0.
\end{equation}
This equation is solved by
\begin{equation}
B(y)=-\frac{1}{8} \ln (4 y + b)^2.
\end{equation}
Here, since $y$ ranges from $0$ to $\infty$, it is mandatory to
assume that $b>0$, so that the 6D space is regular everywhere.
Furthermore,
\begin{equation}
\tilde{R}_6 = \frac{20(1-r^2) \, e^{-2B}}{r^2},\quad
\tilde{R}_{ijkl} \tilde{R}^{ijkl} =\left(\frac{136}{r^4}-\frac{80
b}{r^2}+40\right) e^{-4B},
\end{equation}
which both are regular at $y=0$. Since $0\le y < \infty$, we have
$$
0 < r^2 < \infty,
$$ which includes all three possibilities that $0<r^2<1$ (positively curved 6D space),
$r^2=1$ (Ricci flat 6D space) and $1< r^2 < \infty $ (negatively
curved 6D space).

By evaluating flux contributions to the 10D supergravity action,
we explicitly find
\begin{eqnarray}
S_{\rm eff} &=& \frac{M_{10}^2}{2} {\rm Vol} (X_5) r^5 \int dy
\int \sqrt{-\hat{g}_{4}}\, d^4{x} \Big[
\frac{1}{(4y+b)^{5/2}}\left( e^{-4A} \hat{R}_4 +
\tilde{R}_6\right) + 2A^{\prime\prime}- 8 {A^\prime}^2
\nonumber \\
&{}& \qquad -\frac{c_1^2}{(4y+b)\,r^6}\, e^{4A}
 - \frac{ (4y+b)}{r^4} \, e^{4A} (\partial_y f)^2
 - \frac{(c_0+2c_1 f(y))^2}{2 r^{10}}\, e^{8A}\Big].
\end{eqnarray}
Indeed, the second order (nonlinear) equations of motion derived
from an effective 4D lagrangian, but ignoring the second
derivative terms, such as $A^{\prime\prime}(y)$, or
$z^{\prime\prime}$ in the notation of~\cite{Buchel-etal}, do not
necessarily solve the full system of 10D supergravity equations.
We shall therefore consider the original system of 10D Einstein
equations, which reduce to~\footnote{In our case, it is sufficient
to consider only the trace parts of 10D Einstein equations.}
\begin{eqnarray}
e^{-4A} \hat{R_4} - 4 (4y+b)^{5/2}\, A^{\prime\prime} = -
(4y+b)^{3/2} e^{4A} \left(  \frac{c_1^2}{r^6} +
\frac{(4y+b)^2}{r^4} (\partial_y f)^2
 + \frac{{\cal F}^2 (4y+b)}{2r^{10}} \,
 e^{4A}\right),\label{sugra-eq1}\nonumber \\
\end{eqnarray}
\begin{eqnarray}
&& 20 \sqrt{(4y+b)} \, \left(\frac{1-r^2}{r^2}\right)
+(4y+b)^{5/2}
\left(6A^{\prime\prime}-8 {A^\prime}^2\right) \nonumber \\
&& \qquad \quad ~~~~~~~~~~~~~~ = (4y+b)^{3/2} \, e^{4A}
\left(\frac{3}{2} \frac{c_1^2}{r^6} + \frac{3(4y+b)^2}{2r^4}
(\partial_y f)^2 + \frac{{\cal F}^2 (4y+b)}{2\, r^{10}}\, e^{4A}
\right),\label{sugra-eq2}\nonumber \\
\end{eqnarray}
where ${\cal F}\equiv c_0+ 2 c_1 f(y)$. The problem simplifies a
lot in the case $\hat{R}_4=0$ and $r^2=1$. This case has been
extensively studied in the literature, see, for
example,~\cite{GKP,Buchel-etal}.

For $0<r^2<1$ and $r^2=1$, which correspond, respectively, to the
positively curved and Ricci flat 6D spaces, there is no large $r$
limit solution. As a result the flux contributions may not be
negligible even when one takes a large $y$ limit. However, in the
case of a negatively curved 6D space, the 5-form flux
contributions can be small in the large $r$ limit. Moreover, as
with standard compact hyperbolic manifolds, the 6D curvature and
Kaluza-Klein mass gap may be fixed by two physical
parameters~\footnote{See, for example, ref.~\cite{SCPark} for a
discussion related to phenomenological advantages of using
hyperbolic extra dimensions.}. The freedom to allow $r^2$ in the
range $1< r^2 < \infty$, or a negatively curved 6D space, is an
important difference with respect to the case of the sphere or
Ricci flat 6D space, where the internal volume and Kaluza-Klein
mass gap are generally fixed in terms one parameter, the radius of
${\cal M}=X_6$.

\subsection{Vacuum case}

In the absence of fluxes, i.e. in the absence of terms on the
right-hand side of equations~(\ref{sugra-eq1}) and
(\ref{sugra-eq2}), the system of equations are explicitly solved
when~\footnote{With $r^2=1$ (and hence $\tilde{R}_6=0$), which are
the original choice made by Kachru el al.~\cite{GKP}, there exists
only a trivial solution with $A(y)={\rm const}$ and $\hat{R}_4=0$.
The non-existence of a 4D de Sitter within the framework
of~\cite{GKP} and in many follow up works was just due to the use
of some oversimplified 10D metric Ans\"atze.}
\begin{equation}
r^2=2, \quad \hat{R}_4= 8 \,e^{4 a_1}, \quad A(y)=-\frac{1}{8} \ln
(4y+b) +a_1.
\end{equation}
The explicit 10D metric solution is given by
\begin{equation}
ds_{10}^2 = e^{2A(y)}\Big[ \hat{g}_{\mu\nu} dx^\mu dx^\nu + 2
e^{-4 a_1} \left(\frac{dy^2}{2(4y+b)^2} + dX_5^2\right)\Big].
\end{equation}
In this case, the extra dimensions are hyperbolic (or negatively
curved). Here, the y coordinate varies from $0$ to $\infty$. Of
course, the 6D volume (without the effect of warping), which is
given by $V_6 \sim \int \sqrt{g_6} d^6 y \sim {\rm Vol}(X_5) \int
(4y+b)^{-1} \propto \ln(4y+b)$ grows logarithmically with $y$.
However, when we take into account the effect of warp factor (or
warping of extra dimensions), then the 6D warped volume, which is
given by
$$
V_6^{{\rm warped}} \sim {\rm Vol}(X_5) \int_0^\infty
\frac{dy}{(4y+b)^2}\propto {\rm Vol}(X_5)
\left[\frac{-1}{4y+b}\right]_0^{\infty} =\frac{{\rm Vol}(X_5)}{b},
$$ is finite. The solution therefore leads to a finite 4D
effective Planck mass.

\subsection{Effects of fluxes}

In the presence of both 3-form and 5-form fields, we may analyse
the system of 10D supergravity equations by imposing the
condition~\footnote{This relationship is typically obtained in
type II supergravity by taking a variation of 10D dilaton $\Phi$
and then by taking the limit where $\Phi\to {\rm const}$.}
\begin{equation}
g_s^2 F_3^2 = H_3^2,
\end{equation}
where $g_s$ is string coupling constant. We then find
\begin{equation}
f(y)= \sqrt{ \frac{g_s^2 c_1^2}{16 r^2} } \, \ln (4y+b)+
\frac{f_0}{|r|},
\end{equation}
where $f_0$ is a constant. We now have
\begin{eqnarray}
&& R^{(4)}= e^{-2A}\hat{R_4} - 4 (4y+b)^{5/2}\, A^{\prime\prime} e^{2A} \nonumber \\
&& \qquad ~~~~~~~~~~~~~~~~~  = - f_1 (4y+b)^{3/2} \, e^{6A} -
\left(f_2 +f_3 \ln (4y+b)\right)^2 (4y+b)^{5/2} e^{10
A},\label{flux-sugra1}
\end{eqnarray}
\begin{eqnarray}
&& R^{(6)}= 20 \sqrt{(4y+b)} \left(\frac{1-r^2}{r^2}\right) e^{2A}
+(4y+b)^{5/2}
\left(6A^{\prime\prime}-8 {A^\prime}^2\right) e^{2A} \nonumber \\
&& \qquad ~~~~~~~~~~~~~~~~ = \frac{3 f_1}{2} (4y+b)^{3/2} \,
e^{6A}
 + \left(f_2 +f_3 \ln (4y+b)\right)^2 (4y+b)^{5/2} e^{10 A}.\label{flux-sugra2}
\end{eqnarray}
The coefficients like $f_1$, $f_2$ and $f_3$ are not fully
determined, except that
\begin{equation}
f_1=\frac{c_1^2}{r^6} \left(1+ g_s^2 \right)>0, \qquad
f_3=\frac{c_1^2 g_s}{2 r^6}>0,
\end{equation}
while the coefficient $f_2\equiv (c_0 r+2c_1 f_0)/r^6$ can take
any values. Our aim here is not to solve the equations
(\ref{flux-sugra1})-(\ref{flux-sugra2}) exactly, which is anyway
not possible ~\footnote{Indeed, the choice as $\beta=-\alpha$ and
$C=5B + {\rm const}$ does not help us to analytically solve the
10D supergravity equations except when the flux contributions
become negligible or in the limit $\hat{R}_4 \to 0$, in the latter
case one may supplement the full system of second order
(nonlinear) equations with BPS saturated first-order equations as
in Klebanov-Strassler work~\cite{KS} (see
also~\cite{GKP,Buchel-etal}), but it nonetheless assists us to
simplify the system of field equations.} but to study the possible
effects of flux terms on the 4D scalar curvature $\hat{R}_4$ and
also on $R^{(6)}$.

The 3-form flux scales as the energy source for the curvature when
$$ e^{4A} \equiv \frac{e^{4 a_1}}{4y+b} .$$
This leads to
\begin{equation}
R^{(6)}= \frac{4(5-r^2)}{r^2}\, e^{4 a_1} = \frac{3 f_1}{2} e^{6
a_1} + \left(f_1+ f_3 \ln (4y+b)\right)^2 e^{10 a_1}.
\end{equation}
We are now required to take $r^2 < 5$. This includes the
possibility of having a negatively curved 6D space, i.e. $r^2>1$.
From the expression of $R^{(4)}$ we can see that $\hat{R}_4>0$
provided that
$$ 16 - f_1 \,e^{4 a_1} - (f_2+ f_3 \ln (4y+b))^2 e^{8 a_1}>0. $$
This can be satisfied by taking small $f_i$s ($i=1,2,3$) or that
$a_1 \ll 0$~\footnote{In this case, one needs to have a relatively
large volume for the extra dimension, or a reasonably small value
of $c_1^2$, since otherwise $\hat{R}_4$ will change its sign from
positive to negative as $y\to \infty$.}.

From equations~(\ref{flux-sugra1})-(\ref{flux-sugra2}), we can see
that a sufficiently long period of 4D de Sitter expansion becomes
possible when the warp factor varies as
\begin{equation}\label{sol-warp}
e^{4A} \equiv \frac{e^{4a\Z1}} {(4y+b)^{\gamma}} \qquad (\gamma>1)
\end{equation}
or more rapidly. In such a case the effect of fluxes fall off more
rapidly as compared to the energy sources for the curvature. For
illustration, take $\gamma=3/2$, so that
\begin{equation}
{R}^{(6)}= \frac{2(10-r^2)}{r^2 (4y+b)^{1/4}}\,e^{2a_1}.
\end{equation}
The positivity of $R^{(6)}$ now requires that $r^2 < 10$.
Furthermore
$$
\hat{R}_4 \to \frac{96\,e^{4 a_1}}{(4y+b)} - f_1 \,
\frac{e^{8a_1}}{(4y+b)^{3/2}} -\frac{(f_2+ f_3 \ln
(4y+b))^2}{(4y+b)^2}\, e^{12 a_1}.$$ Qualitatively, we can have
three branches of solutions:
\begin{eqnarray}
0< y < y_c : && \quad \hat{R}_4  > 0, \nonumber \\
y_c < y < y_2 : && \quad \hat{R}_4 < 0, \nonumber \\
y  \to \infty : && \quad \hat{R_4} \sim \frac{e^{4a_1}}{y} \left(
24 - f_1 \frac{e^{4a_1}}{8 y^{1/2}} + \cdots\right)>0. \nonumber
\end{eqnarray}
Here, $y_c$ is some critical value of $y$, for which the energy
sources of the curvature and 3-form gauges fields become
comparable and hence $\hat{R}_4 \sim 0$. That is, in the presence
of fluxes, there could arise an intermediate branch where
$\hat{R}_4<0$ is possible, thereby generating an epoch of
decelerating expansion between two de Sitter
expansions~\footnote{The second phase of an accelerating expansion
is not purely de Sitter since $\hat{R}_{4}$ is no longer a
constant.}. It would be very useful to find an explicit solution
of 10D supergravity equations realizing such a behaviour.

\section{A model with compact extra dimensions}

In this section, which may be read independently of the above
sections, we will briefly discuss an alternative scenario where
the fifth dimension is compact (similar to that in the RS 2-brane
model). For simplicity, we will drop all source terms (or fluxes).

It is not difficult to check that the following 11-dimensional
metric ansatz
\begin{equation}
ds_{11}^2 =e^{K(y)} \left(-dt^2 + e^{2m_4 t} d{\bf x}_3^2 +
\lambda^2 \tanh^2{y} \,dy^2 +\rho^2
d\Omega_6^2\right),\label{11d-metric}
\end{equation}
where $d\Omega_6^2$ denotes the metric of a standard
six-sphere~\footnote{ One could in principle replace $S^{m}$ by
Einstein-Sasaki spaces $(S^{m-n}\times S^{n-1}) \rtimes S^{1}$ or
some other compact manifolds, in a way similar to the 10D case
where gravity solutions with $X_5=S^5$ are physically equivalent
to that with $X_5=T^{1,1}$, at least, in pure Einstein gravity
(i.e. without source terms).}, explicitly solves all of the 11D
Einstein field equations when
\begin{equation}
K(y)=  \pm \sqrt{\frac{20\lambda^2}{9 \rho^2}} \ln\cosh{y}+K_0,
\qquad m_4= \sqrt{\frac{5}{3\rho^2}}.\label{11d-sol}
\end{equation}
Here we take the negative sign~\footnote{Only the negative sign in
this solution leads to a finite 4D effective Planck mass once the
orbifold symmetry around $y=[0,\pi]$ is relaxed. As in the 10D
case, we can get a more general solution by replacing $\cosh{y}$
by $\cosh{y}+\epsilon\sinh{y}$ and $\tanh{y}$ by
$(\epsilon+\tanh{y})/(1+\epsilon \tanh{y})$.}. As the $y$
coordinate is assumed to be a closed cycle, we can write the total
action as
\begin{equation}
S=\frac{M_{11}^2}{2} \int d^{11} x \sqrt{-g_{11}}\,R_{11} +
\int_{M^4 \times {\cal M}} \sqrt{-g_{b1}} \left(-\tau_{b1}\right)
+ \int_{M^4 \times {\cal M}} \sqrt{-g_{b2}}
\left(-\tau_{b2}\right),
\end{equation}
where $\tau_{b1}$, $\tau_{b2}$ denote the brane tensions
corresponding to the two p-branes $b1$ and $b2$, which may be
placed at orbifold fixed points $y=0$ and $y=\pi$. Here, we are
assuming the existence of something like a D9-brane, i.e. $p=9$,
so each brane has 10 dimensions. The 6 spatial directions of
p-brane extend in the internal manifold $X_6$, while the 4 other
directions (including one time-like) extend along the $x^\mu$
directions. The Horava--Witten `world brane'~\cite{Horava96} is
such a brane.

The solution valid for $-\pi \le y \le \pi$ implies that
\begin{equation}
K^{\prime\prime} + \frac{2\sqrt{5}\lambda}{3\rho\cosh^2{y}} +
\frac{2\sqrt{5}\lambda}{3\rho} \tanh{y} \left(2\delta (y)
-2\delta(y-\pi)\right)=0.
\end{equation}
From the $\mu\nu$ components of the 11D Einstein equations, we
obtain
\begin{eqnarray}
&& \frac{9\,e^{-{K}/{2}}}{\lambda \tanh{y}} \left(K^{\prime\prime}
-\frac{2K^\prime}{\sinh(2y)} + 2 {K^\prime}^2 -
\frac{2\lambda^2\tanh^2{y}}{3}\left( m_4^2 + \frac{15}{3\rho^2}
\right) \right) + \frac{4\tau_{b1}}{M_{11}^2} \delta (y) +
\frac{4\tau_{b2}}{M_{11}^2} \delta(y-\pi)= 0 \nn \\
&& \Rightarrow  \frac{9\,e^{-K/2}}{\lambda \tanh{y}}
\left(K^{\prime\prime} + \frac{2\sqrt{5}\lambda}{3\rho\cosh^2{y}}
\right) + \frac{4\tau_{b1}}{M_{11}^2} \delta (y) +
\frac{4\tau_{b2}}{M_{11}^2} \delta(y-\pi)=0.\label{11d-sol2}
\end{eqnarray}
In the second line above we have substituted the solution
(\ref{11d-sol}). The $ij$ components ($i,j=6, \cdots, 11$) of the
11D Einstein tensor $G_{AB}$ also lead to the same result (cf
(\ref{11d-sol2})).

By comparing the above set of equations, we explicitly get
\begin{equation}
\tau_{b1} = \frac{3\sqrt{5}}{\rho M_{11}^2 }\,e^{-K_0/2}, \qquad
\tau_{b2}=- \frac{3\sqrt{5}}{\rho M_{11}^2} \,e^{-K_0/2}
\exp\left(-\frac{5\lambda}{3\sqrt{2}\rho} \ln \cosh{\pi} \right).
\end{equation}
With a suitable choice of the ratio $\lambda/\rho$, one could
possibly explain the mass hierarchy problem in particle physics,
similar to that in the RS 2-brane model~\cite{RS1}.
From (\ref{11d-metric}), we can see that the warp factor acquires
a low value at $y=\pi$ (compared to that at $y=0$), which is
necessary for the hierarchy solution. The compactification scale
$\rho$, which is proportional to the size of $X_6$, is not
necessarily small within our model. We only require that the ratio
$\lambda/\rho$ becomes much larger than unity, e.g. $\lambda/\rho
\sim {\cal O}(10)$. For instance, with $\lambda/\rho\sim 12 $, we
obtain $\tau_{b_1}/|\tau_{b_2}| \sim {\it e}^{34.7}\sim 10^{15}$.
In the RS 2-brane model, since $p=3$, the brane is extended only
along the $x^\mu$ directions. Here, as we are considering a
p-brane with $p=9$, the brane is extended also along the internal
manifold $X_6$. That means, we are comparing the 9-brane tensions
at the orbifold fixed points $y=0$ and
$y=\pi$~\footnote{Alternatively, one can introduce a combination
of D3 and M5 branes and also covariant combinations of delta
functions and geometric factors necessary to position the branes.
In such a case one is usually required to introduce some local
sources concentrated (or localized) in the manifold $X_6$ in order
to solve 11D Einstein equations at the brane's positions.}.

Next we briefly discuss about phenomenological constraints on the
model. Indeed, the compactification scale $\rho$ must respect some
experimental bounds which assert for a low $\rho$. To be precise,
we shall consider the mass reduction formula (also known as
Gauss's law)
\begin{equation}
M_{\rm Pl}^2 = \frac{M_{(11)}^9\,e^{9K_0/2}}{(2\pi)^7 } \,
V_7^{\rm w}, \qquad V_7^{\rm w}= {\rm Vol}(X_6)\,
\frac{\rho^7}{3\sqrt{5}}\left[1- \left({\rm
sech}\pi\right)^{3\sqrt{5}\lambda/\rho}\right],\end{equation}
which is obtained by considering a dimensional reduction from
$D=11$ to $D=4$. The 7D warped volume is well approximated by
$V_7^{\rm w}\simeq {\rm Vol}(X_6) \rho^7/(3\sqrt{5})$ for
$\lambda/\rho \gtrsim 1$. There arises a strong constraint on
$\rho$, especially, if one wants to tune $\Lambda_4$ to the
present value of the cosmological constant, $\Lambda_4\sim
10^{-120}\,M_{\rm Pl}^2\sim 10^{-55} {\rm cm}^{-2}$, which is
given here by $\Lambda_4 \sim m_4^2 \sim 1/\rho^2 $ (cf
equation~(\ref{11d-sol})). However, when one applies the model to
explain a period of de Sitter expansion in the early universe,
then the constraint on $\rho$ is mild and it can easily be
satisfied. For instance, with $\rho^{-1} \sim 10^{-5}\,M_{\rm
Pl}$, we obtain $M_{11}\sim 5.5 \times 10^{-5}\,M_{\rm Pl}\,
e^{-K_0/2}$. Alternatively, we may demand that $M_{\rm Pl} >
M_{11} \gtrsim 10^{-15}\,M_{\rm Pl}$. Then the bound on $\rho$
turns out to be $0.15\,{\it e}^{K_0/2}\, M_{\rm Pl} > \rho^{-1} >
0.79\times 10^{-20}\, {\it e}^{K_0/2}\,M_{\rm Pl}$. In view of
these bounds, the explicit solutions found in this paper may be
used only to explain a de Sitter expansion in the early universe
rather than a late epoch cosmic acceleration.

\section{Summary and further remarks}

Finding de Sitter solutions through warped compactifications of
type-II supergravities in 10 or 11 dimensions has often been
viewed difficult, except in the cases where one supplements the
gravitational action with some stringy sources that violate
certain positivity conditions. This is more or less the gist of an
argument in several versions of no-go theorem for de Sitter
compactifications of classical supergravity theories, with
physically compact extra dimensions.

In the literature, see, for example,
~\cite{Silverstein2007,Haque2008}, there are arguments about the
existence of (universal) de Sitter solutions at tree-level
supergravity actions in 10 or 11 dimensions, which may involve
only compact extra dimensions. The proofs of existence of 4D de
Sitter solutions in these and several other constructions have
been based on certain assumptions under which part of the (super)
symmetries are broken, giving rise to an effective (scalar field)
potential which allows a local minimum, rather than on exact
solutions to classical field equations. In this paper we have
found a few sufficiently illustrative examples in which
four-dimensional de Sitter solutions arise at a purely classical
level.

The most interesting achievement of this paper is the discovery of
a new class of simple warped vacuum solutions with a 4D de sitter
part and noncompact internal space with still a finite 4D Planck
mass~\footnote{In a recent
 paper~\cite{Ish10e}, it has been shown that
 the class of de Sitter solutions found in this paper also lead to a
 normalizable zero-mass graviton wave-function in four dimensions.}.
It is quite plausible that warped compactifications on Ricci
non-flat spaces, including manifolds whose mean curvature is
negative, lead to the realization of some new
cosmological scenarios in string and supergravity theories. 

\section*{Acknowledgement}

The author is grateful to Kei-ichi Maeda, Shinji Mukohyama, Hideo
Kodama, Nobu Ohta and Thomas Van Riet for useful conversations and
correspondences. This work was supported by the Marsden fund of
the Royal Society of New Zealand.


\end{document}